\documentclass[10pt, twocolumn]{IEEEtran}

\newcommand{\secref}[1]{Sec.~\ref{#1}}
\IEEEoverridecommandlockouts
\usepackage{cite}
\usepackage{amsmath,amssymb,amsfonts}
\usepackage{amsthm}
\usepackage{enumitem}
\usepackage{algorithmic}
\usepackage{graphicx}
\usepackage{textcomp}
\usepackage{xcolor}
\usepackage[]{footmisc}
\usepackage{bbm}

\newcommand{\h}{{\mathrm H}}

\usepackage{cuted}
\usepackage{stfloats}
\usepackage{mathtools,amssymb,lipsum, nccmath}

\newtheorem{example}{Example}
\usepackage{algorithm,algorithmic}
\usepackage{color,soul}
\usepackage{graphicx}
\usepackage{mathrsfs}
\usepackage{aligned-overset}
\usepackage{caption,subcaption}

\begin{document}

\pagenumbering{gobble} 

\title{Interference-Robust Non-Coherent Over-the-Air Computation for Decentralized Optimization}
\author{Nicolò Michelusi,~\IEEEmembership{Senior Member, IEEE}
\thanks{N. Michelusi is with the School of Electrical, Computer and Energy
Engineering, Arizona State University. email: 
nicolo.michelusi@asu.edu.
This research has been funded in part by NSF under grant CNS-$2129015$.}
\vspace{-5mm}
}

\maketitle
\begin{abstract} 
Non-coherent over-the-air (NCOTA) computation enables low-latency and bandwidth-efficient decentralized optimization by exploiting the average energy superposition property of wireless channels. It has
recently been proposed as a powerful tool for executing consensus-based optimization algorithms in fully decentralized systems. A key advantage of NCOTA is that it enables unbiased consensus estimation without channel state information at either transmitters or receivers, requires no transmission scheduling, and scales efficiently to dense network deployments. However, NCOTA is inherently susceptible to external interference, which can bias the consensus estimate and deteriorate the convergence of the underlying decentralized optimization algorithm. In this paper, we propose a novel interference-robust (IR-)NCOTA scheme. The core idea is to apply a coordinated random rotation of the frame of reference across all nodes, and transmit a pseudo-random pilot signal, allowing to transform external interference into a circularly symmetric distribution with zero mean relative to the rotated frame. This ensures that the consensus estimates remain unbiased, preserving the convergence guarantees of the underlying optimization algorithm. 
Through numerical results on a classification task, it is demonstrated that IR-NCOTA exhibits superior performance over the baseline NCOTA algorithm in the presence of external interference.
\end{abstract}
\vspace{-1mm}
\vspace{-2mm}
\section{Introduction}
\label{sec:intro}
 Decentralized optimization and learning arise in various domains such as remote sensing, distributed inference \cite{6494683}, estimation \cite{9224135}, multi-agent coordination \cite{Nedic2018}, and machine learning (ML)  \cite{YANG2019278}. 
 In traditional learning frameworks, data is aggregated in a centralized location where computationally intensive optimization algorithms are executed. 
 Yet, in {emerging verticals} where infrastructure is absent, unreliable, or has been disrupted, such as post-disaster environments \cite{https://doi.org/10.1002/rob.22075}, search and rescue operations, or remote rural regions \cite{9475989}, centralized aggregation may be impractical.
 These limitations require a shift toward decentralized alternatives, in which agents
 such as uncrewed aerial vehicles (UAVs)
  perform sensing, inference, and learning locally, while relying only on  peer-to-peer communications~\cite{8950073}.
Within this context,
this paper aims to solve the optimization problem
\begin{align}
\label{global}
{\mathbf w}^*=\arg\min_{\mathbf w\in\mathbb R^d}\ F(\mathbf{w})\triangleq\frac{1}{N}\sum_{i=1}^Nf_i(\mathbf w) 
\end{align}
among $N$ wirelessly-connected nodes,
where $f_i(\mathbf w)$ is the local function of node $i$, known to $i$ alone,
and $\mathbf w$ is a $d$-dimensional parameter vector.
For instance,  $f_i$ may represent the empirical loss
 based on the local dataset of node $i$, and $F(\mathbf{w})$ is the empirical loss over the global dataset.

  \begin{figure}
     \centering
         \includegraphics[width = .7\linewidth]{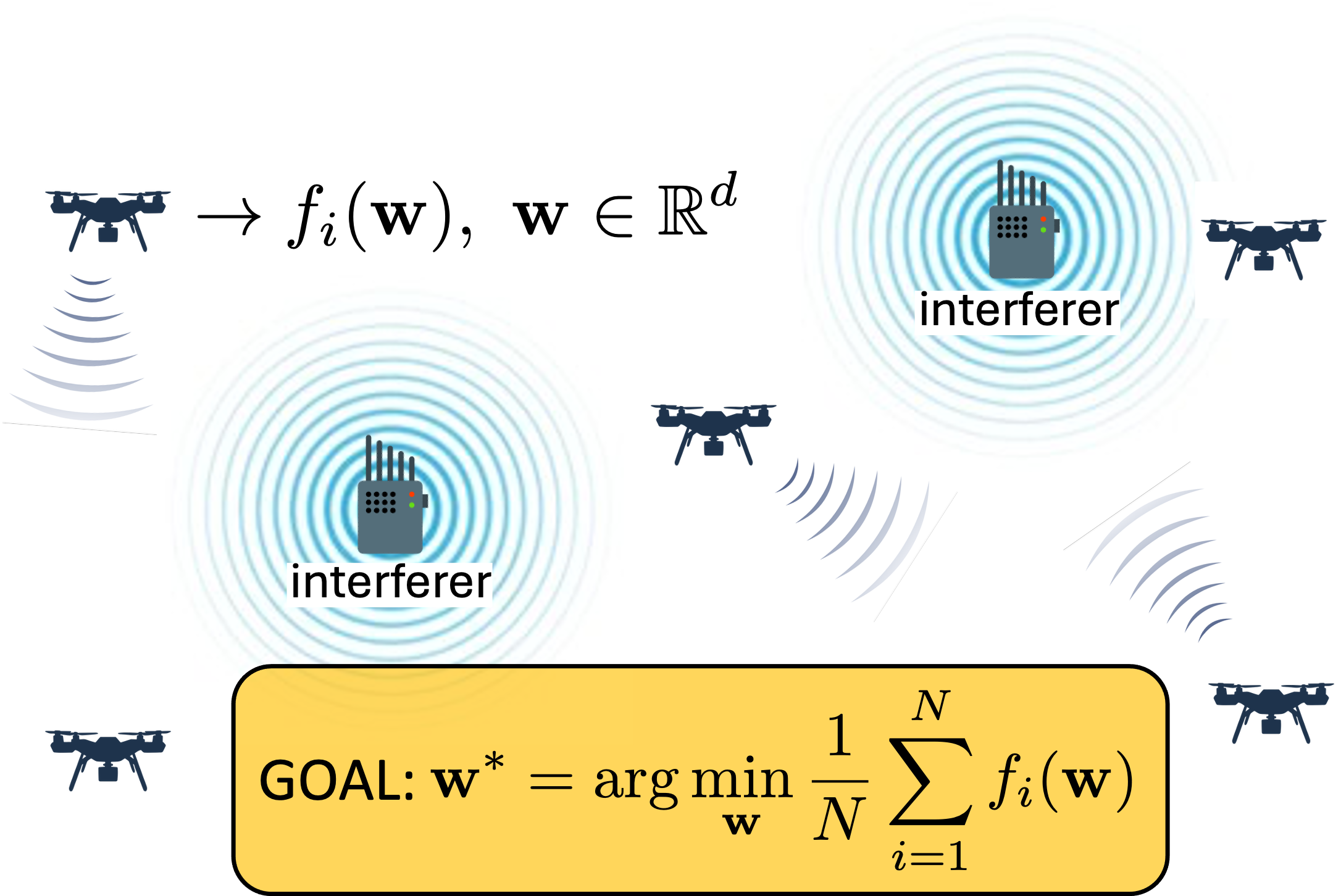}
         \caption{Example of a swarm of UAVs collaboratively solving a decentralized optimization problem, with two  interferers generating unwanted interference.}
          \vspace{-6mm}
\label{fig:sigmo}
\end{figure}
 
A renowned algorithm to solve~\eqref{global} in decentralized settings is \emph{Decentralized Gradient Descent} (DGD)~\cite{Nedic2009grad,Yuan2016}. In DGD, each node $i$ iteratively updates its local parameter vector $\mathbf w_i$ through consensus and local gradient descent. However, implementing DGD over wireless channels poses significant challenges due to noise, fading, interference, and the need for multi-access coordination~\cite{10447686,10279097,9563232,9517780,9322286,9772390,9716792,9838891}.  

In a wireless implementation of DGD~\cite{10947567,10680589}, each node updates its local parameter vector at iteration $k$ as
\begin{align}
\mathbf w_{i,k+1}
= \mathbf w_{i,k} + \gamma_k \hat{\mathbf d}_{i,k} - \eta_k \nabla f_i(\mathbf w_{i,k}),
\tag{DGD}
\label{ddd}
\end{align}
where $-\eta_k \nabla f_i(\mathbf w_{i,k})$ represents a local gradient descent step controlled by the learning stepsize $\eta_k$, and $\hat{\mathbf d}_{i,k}$ is an estimate of the \emph{disagreement signal}, defined as
\begin{align}
\label{diklap}
\mathbf d_{i,k} = -\sum_{j=1}^N \ell_{ij} \mathbf w_{j,k}
= -\sum_{j \neq i} \ell_{ij} (\mathbf w_{j,k} - \mathbf w_{i,k}),
\end{align}
where $\{\ell_{ij}\}$ denote symmetric Laplacian weights, satisfying $\ell_{ij}{=}\ell_{ji}{\leq}0$ for all $i{\neq}j$ and $\ell_{ii}{=}-\sum_{j \neq i} \ell_{ij}$. The parameter $\gamma_k{>}0$ is a consensus step size that mitigates the effect of errors in the disagreement signal estimation caused by imperfect wireless communications (e.g., noise, fading, or interference).  
 Since $\sum_j \ell_{ij}{=}0$, if all local parameter vectors $\mathbf w_{j,k}$ take a common value $\bar{\mathbf w}$, then $\mathbf d_{i,k}{=}\mathbf 0$. 
Thus, $\mathbf d_{i,k}$ quantifies the degree of disagreement of $\{\mathbf w_{i,k}\}$ across the network.

An estimate of $\mathbf d_{i,k}$ is discussed in Secs. \ref{bckr} and \ref{sysmo}. In essence, DGD combines a correction step based on this estimate ($\hat{\mathbf d}_{i,k}$), which promotes agreement and reduces model discrepancies, with a local gradient descent step that drives convergence toward optimality. Iterating these steps steers the system toward consensus and global optimality.


In our prior work \cite{10947567,10680589}, we demonstrated that, for strongly convex $f_i$, the noisy DGD algorithm converges to the global optimum $\mathbf w^*$ with an error $
\frac{1}{N}\sum_i\mathbb{E}[\|\mathbf w_{i,k}{-}\mathbf w^*\|^2] = \mathcal{O}(1/\sqrt{k})$, for a suitable choice of  $\eta_k,\gamma_k$.
However, this convergence result relies on \emph{unbiasedness} of the disagreement signal estimate, i.e., $\mathbb{E}[\hat{\mathbf d}_{i,k}|\mathbf w_{j,k}, \forall j]{=}\mathbf d_{i,k}$, as it ensures that fluctuations of the estimate $\hat{\mathbf d}_{i,k}$ around its expected value $\mathbf d_{i,k}$ are averaged out over DGD iterations. 
Conversely, if this unbiasedness condition is not satisfied, 
a distortion is introduced that accumulates over time, preventing convergence.

Recently, non-coherent over-the-air computation (NCOTA) has been proposed in~\cite{10680589} to estimate the disagreement signal over wireless channels affected by noise and fading. In this scheme, each local parameter vector $\mathbf w_{i,k}$ controls the energy of the transmitted signal. Nodes transmit in a randomized and simultaneous fashion, and their signals naturally superimpose over the wireless medium, producing an \emph{unbiased estimate} of the disagreement signal by exploiting the \emph{average energy superposition} property of wireless channels. This enables unbiased estimation of $\mathbf d_{i,k}$ without requiring topology information, channel state information at the transmitters or receivers, or transmission scheduling, at the cost of additional noise in the disagreement signal estimation due to channel impairments such as noise, fading, and energy fluctuations. Numerical results in~\cite{10680589} demonstrate that NCOTA outperforms implementations based on orthogonal transmissions \cite{8786146} and conventional over-the-air computation \cite{8870236,9563232},
 particularly as the number of agents increases.

However, the unbiasedness of NCOTA no longer holds under external interference, as interference energy accumulates at the receivers, introducing an uncontrolled drift in the DGD updates that undermines convergence. In this paper, we propose a novel \emph{Interference-Robust} (IR) NCOTA scheme that preserves the effectiveness of NCOTA even under external interference.
The key idea is to leverage coordinated randomness across the network to \emph{scramble} the interference signals.
 Specifically, we introduce two key mechanisms: (1) a coordinated random rotation of the frame of reference, which maps energy values to different reference orientations so that, when averaged over the pseudo-random rotations, the additive interference contribution appears zero-mean; and (2) a pseudo-random pilot sequence, used to estimate the aggregate channel gains (self Laplacian weights $\ell_{ii}$) while ensuring that the interference contribution remains zero-mean. We show that IR-NCOTA produces an unbiased estimate of the disagreement signal, enabling direct application of the convergence results established in~\cite{10680589}.

The remainder of this paper is structured as follows.
\secref{bckr} provides a background of NCOTA.
\secref{sysmo} presents the proposed algorithm, and demonstrates its unbiasedness properties.
\secref{numres} provides numerical results.
Finally, \secref{conclu} provides concluding remarks.
 We refer to our prior work \cite{10680589} for an in-depth
discussion of NCOTA and its convergence properties.

\noindent\underline{Notation:} 
We use boldface letters for vectors (e.g., $\mathbf a \in \mathbb C^d$) and non-boldface letters for scalars (e.g., $a \in \mathbb C$). 
For a (column) vector $\mathbf a$, we define:
its $m$th element as $[\mathbf a]_m$;
its conjugate transpose as $\mathbf a^\h$; its transpose as $\mathbf a^\top$;
its Euclidean norm as $\|\mathbf a\| = \sqrt{\mathbf a^\h \mathbf a}$; and its $\ell_1$ norm as $\|\mathbf a\|_1 = \sum_m |[\mathbf a]_m|$. 
 For a scalar $a \in \mathbb C$, we denote its complex conjugate as $a^*$. 
We use $\mathbf 1$ and $\mathbf 0$ to denote the all-ones and all-zeros vectors, respectively,
$\mathbf e_m$ for the $m$th standard basis vector,
 and $\mathbf I$ to denote the identity matrix (with dimension clear from context). 
Finally, $\mathcal{CN}(\mathbf b, \Sigma)$ denotes a circularly symmetric complex Gaussian random vector with mean $\mathbf b$ and covariance matrix $\Sigma$.


\vspace{-3mm}
\section{Background of NCOTA \cite{10680589}}
\label{bckr}

We assume that $\|\mathbf w^*\|\leq r$ for some known $r$, hence the optimization is restricted to the $d$-dimensional sphere $\mathcal W \equiv \{\mathbf w \in \mathbb R^d : \|\mathbf w\| \leq r\}$. This guarantees that signals remain bounded during communication, thereby ensuring practical energy constraints.
For notational convenience, we omit the dependence on the iteration index $k$.
NCOTA estimates the disagreement signal over wireless channels as follows \cite{10680589}. 

\underline{Energy-based encoding:} First, the local parameter vector $\mathbf w_i\in\mathcal W$
is expressed as the convex combination
 \begin{align}
\label{convcomb0}
{\mathbf w}_i=\sum_{m=1}^{M}[\mathbf p_{i}]_m\mathbf z_{m},
\end{align}
where $\{\mathbf z_1,\dots,\mathbf z_{M}\}$ is a set of $M\triangleq 2d+1$ codewords defined as
$\mathbf z_{M}=\mathbf 0$ and, for $m=1,\dots,d$,
\begin{align}
\label{CpZ}
\mathbf z_m=\sqrt{d}r\mathbf e_m,\ \mathbf z_{d+m}=-\sqrt{d}r\mathbf e_m.
\end{align}
whereas $\mathbf p_{i}$ is a set of non-negative coefficients defined as
 $[\mathbf p_{i}]_{M}{=}1{-}\frac{1}{\sqrt{d}r}\Vert\mathbf w_{i}\Vert_1$, and, for $m{=}1,\dots,d$,
\begin{align}
\label{pim}
[\mathbf p_{i}]_m{=}\frac{1}{\sqrt{d}r}([{\mathbf w}_i]_m)^+,[\mathbf p_{i}]_{d+m}{=}\frac{1}{\sqrt{d}r}(-[{\mathbf w}_i]_m)^+,
\end{align}
where $(\cdot)^+=\max\{\cdot,0\}$.
With this definition, and using the fact that $a = (a)^+ - (-a)^+$, it is straightforward to verify that \eqref{convcomb0} holds. Furthermore, since $\|\mathbf w_i\|_1\leq \sqrt{d}\|\mathbf w_i\| \leq \sqrt{d}r$ (by the Cauchy-Schwarz inequality) and $(\cdot)^+ \geq 0$, it follows that $\mathbf p_i\geq 0$ and $\mathbf 1^\top\mathbf p_i=1$. We compactly denote the encoding scheme applied by each node through \eqref{pim} as
\begin{align}
\label{pifun}
\mathbf p_i \triangleq \mathcal{P}(\mathbf w_i).
\end{align}
Since $\mathbf p_i \geq 0$, this representation is suitable for fully decentralized energy-based transmission, unlike the signal $\mathbf w_i$, which may contain negative elements. Each transmitter thus generates its transmit signal as
\begin{align}
\label{xenc}
\mathbf x_i = \sqrt{EM}\,\sqrt{\mathbf p_i},
\end{align}
where $\sqrt{\cdot}$ is element-wise, so that $\mathbf p_i$ controls sample energy. Since $\mathbf 1^\top \mathbf p_i = 1$, the energy per sample is $\|\mathbf x_i\|^2 /M = E$.

\underline{Randomized transmission:} 
To satisfy half-duplex constraints, we adopt a probabilistic transmission strategy: with probability $p_{tx}$, node $i$ operates as a transmitter (indicated by $\chi_i = 0$); otherwise, it operates as a receiver (i.e., $\chi_i = 1$). These transmission decisions are i.i.d. across nodes and iterations, enabling a fully decentralized implementation.

\underline{Received signal:} 
Assuming Rayleigh flat-fading channels $h_{ij} \sim \mathcal{CN}(0, \Lambda_{ij})$ between transmitter $j$ and receiver $i$,\footnote{Our paper \cite{10680589} discusses extensions to a broad class of frequency-selective channels.} with average channel gain $\Lambda_{ij}$, independent across node pairs $i,j$, the received signal at receiver $i$ is
\begin{align}
\mathbf y_i = \sum_{j:\chi_j=0} h_{ij}\mathbf x_j + \mathbf n_i,
\label{sigmod}
\end{align}
where $\mathbf n_i \sim \mathcal{CN}(\mathbf 0, N_0 \mathbf I)$ denotes Gaussian noise with power spectral density $N_0$. We will include interference in \secref{sysmo}.

\underline{Disagreement signal estimation:} 
Upon receiving the signal $\mathbf y_i$, each node computes the received sample energy as
\begin{align}
\label{rim}
[\mathbf r_i]_m{=}\chi_i \frac{|[\mathbf y_i]_m|^2 - N_0}{(1 - p_{tx}) p_{tx} E \cdot M}, \  \forall m = 1, \dots, M,
\end{align}
and estimates the disagreement signal as
\begin{align}
\label{dik}
\hat{\mathbf d}_i = \sum_{m=1}^{M} [\mathbf r_i]_m (\mathbf z_m - \mathbf w_i).
\end{align}
Note that $\mathbf r_i{=}\mathbf 0$ and $\hat{\mathbf d}_i{=}\mathbf 0$ for nodes operating as transmitters in the current iteration.

Taking the expectation of $\mathbf r_{i}$ with respect to Gaussian noise $\mathbf n_i$, Rayleigh fading channels $h_{ij}$,
 the randomized transmit/receive decisions $\chi_j\in\{0,1\}$;
and using the energy-based signal encoding \eqref{xenc}, it can be shown that
$$
\mathbb E[\mathbf r_{i}|\mathbf w_j,\forall j]=\sum_{j\neq i} \Lambda_{ij}\mathbf p_{j}.
$$
Then, using \eqref{dik}, the convex combination structure \eqref{convcomb0}, and the fact that $\sum_{m=1}^{M}[\mathbf p_{j}]_m=1$, we obtain
\begin{align}
\nonumber
&\mathbb E[\hat{\mathbf d}_{i}|\mathbf w_j,\forall j]=
\sum_{j\neq i} \Lambda_{ij}\sum_{m=1}^{M}[\mathbf p_{j}]_m(\mathbf z_m-{\mathbf w}_{i})
\\&
=
\sum_{j\neq i} \Lambda_{ij}({\mathbf w}_{j}-{\mathbf w}_{i})\triangleq \mathbf d_i.
\label{targetd}
\end{align}
Comparing this result with \eqref{diklap}, we conclude that
$\hat{\mathbf d}_{i}$ is an unbiased estimate of the disagreement signal, with Laplacian weights $\ell_{ij}{=}{-}\Lambda_{ij}$ given by the average channel gains.

This unbiasedness property is critical to achieve the convergence properties of the DGD algorithm, established in \cite{10680589}. In the next section, we consider a more general signal model that includes external interference. We will show that, under external interference, the estimate given in \eqref{dik} becomes biased, introducing a drift in the DGD updates that accumulates over time and leads to a loss of convergence. We will then propose a novel interference-robust NCOTA scheme that recovers the unbiasedness condition.

\section{Signal Model and IR-NCOTA}
\label{sysmo}

We consider a more general model than \eqref{sigmod}, including external interference, as illustrated in Fig. \ref{fig:sigmo}. Specifically, we model $\mathbf n_i$ as a combined noise-plus-interference term. 
We impose no specific distributional assumptions on $\mathbf{n}_i$, other than a bounded second-order moment and the requirement that it be uncorrelated with the Rayleigh channel coefficients, i.e.,
$
\mathbb{E}\!\left[h_{ij}^* \mathbf{n}_i\right] = \mathbf{0}.
$
The term $\mathbf{n}_i$ may have a non-zero mean, exhibit correlation or non-stationarity, be non-Gaussian, and may even depend on the local optimization signals $\mathbf{w}_j$, for example in the presence of adversarial devices. This model captures a wide range of interference sources, including jamming and adversarial transmitters \cite{6848224}.

Under such more general model, NCOTA fails to compute an \emph{unbiased} estimate of the disagreement signal, as shown in the following example.
\begin{example}
Consider the noise-plus-interference model
\[
\mathbf n_i = g_i\sqrt{E\cdot M}\,[1,0,0,\dots,0]^\top + \mathcal{CN}(\mathbf 0, N_0 \mathbf I),
\]
where $g_i \sim \mathcal{CN}(0, \Gamma_i)$ represents the Rayleigh fading channel between the interferer and node $i$,
generating interference only on the first received sample. Then, under NCOTA, the expectation of $\mathbf r_i$ yields
\[
\mathbb{E}[\mathbf r_i | \mathbf w_j, \forall j]
= \sum_{j \neq i} \Lambda_{ij} \mathbf p_j
+ \frac{\Gamma_i}{p_{tx}}\,[1,0,0,\dots,0]^\top,
\]
and consequently, after applying \eqref{dik} and similarly to \eqref{targetd},
\[
\mathbb{E}[\hat{\mathbf d}_i | \mathbf w_j, \forall j]
= \sum_{j \neq i} \Lambda_{ij} (\mathbf w_j - \mathbf w_i)
+ \frac{\Gamma_i}{p_{tx}} (\mathbf z_1 - \mathbf w_i),
\]
containing both the desired disagreement signal, and a distortion term.
This distortion contains both a fixed bias term proportional to $\mathbf z_1$ and a signal-dependent drift term proportional to $\mathbf w_i$, which accumulate over time and degrade the convergence performance of DGD.
\end{example}

The primary source of performance degradation is that NCOTA relies on the \emph{energy superposition} property of wireless channels.
Specifically, the term "$-N_0$" in~\eqref{rim} compensates for the energy contribution of additive noise.
In contrast, in the presence of an unknown interference source, NCOTA inevitably accumulates energy from that interference.
If the interference structure is unknown, this accumulated interference energy cannot be compensated for, unlike the noise term "$-N_0$" in~\eqref{rim}.

Next, we enhance NCOTA by introducing two key techniques to achieve robustness against interference. We refer to the resulting scheme as \emph{interference-robust} (IR-) NCOTA.
We start by decomposing the target disagreement signal in \eqref{targetd} into two components:
\begin{align}
\mathbf d_i=\underbrace{\sum_{j \neq i} \Lambda_{ij} \mathbf w_j}_{\text{(a)}\ \triangleq \bar {\mathbf s}_i}
-\underbrace{\Big( \sum_{j \neq i} \Lambda_{ij} \Big)}_{\text{(b)}\ \triangleq \bar\Lambda_{i}}\mathbf w_i.
\label{didecomp}
\end{align}
The first term represents (a) 
\emph{weighted sum of the local parameter vectors} across the network, weighted by their respective average channel gains.
We denote this term as $\bar {\mathbf s}_i$ for node $i$.
 The second term (b) 
scales the local parameter vector $\mathbf w_i$ (known only to node $i$) by the \emph{sum of the average channel gains incoming into node $i$}, denoted as $\bar\Lambda_{i}$. We aim to estimate these two terms separately.

\subsubsection{Estimation of the weighted sum of the local parameter vectors, $\bar {\mathbf s}_i$}

In the baseline NCOTA algorithm, during the energy encoding process, the local optimization signals are expressed with respect to a fixed and common frame of reference defined by the codewords $\{\mathbf z_1, \dots, \mathbf z_{M}\}$, as shown in \eqref{pim}. The resulting vectors $\mathbf p_i$ of non-negative coefficients are then used to scale the sample energy of the transmitted signals. 
This creates a vulnerability, as the transmitted energy becomes susceptible to interference that accumulates coherently over time with respect to the \emph{fixed} frame of reference.

To estimate $\bar{\mathbf{s}}_i$, we introduce a key mechanism that
\emph{scrambles} the distortion caused by the energy of the interference, rendering
the interference contribution to the disagreement signal estimation
 a zero-mean process. To this end, each transmitter applies a \emph{pseudo-random frame-of-reference rotation} to its local parameter vector. 
This is achieved by applying a coordinated random unitary transformation to each local parameter vector $\mathbf w_i$:
\[
\tilde{\mathbf w}_i = \mathbf U\cdot \mathbf w_i,
\]
where $\mathbf U\in\mathbb R^{d\times d}$ satisfies $\mathbf U^\top \mathbf U = \mathbf I$, $\mathbb{E}[\mathbf U] = \mathbf 0$, i.i.d. across iterations, and is common to all nodes in the network (e.g., generated using a pseudo-random sequence with a common seed). 
An example  is a random sign flip $\mathbf U = s \mathbf I$, where $s{\in}\{+1, -1\}$ with $\mathbb{P}(s{=}1) = \mathbb{P}(s{=}-1) = 1/2$.
Importantly, $\mathbf U$ is statistically independent of the interference signal $\mathbf n_i$.

Since $\mathcal{W}$ is a sphere centered at $\mathbf 0$, the rotated vector $\tilde{\mathbf w}_i$ also belongs to $\mathcal{W}$, allowing it to be encoded using the same energy encoding scheme described in \secref{bckr}.
Therefore, each transmitting node maps its rotated signal as in \eqref{pifun} to
\[
\tilde{\mathbf p}_i = \mathcal{P}(\tilde{\mathbf w}_i),
\]
so that
\begin{align}
\label{convcomb}
\tilde{\mathbf w}_i = \sum_{m=1}^{M} [\tilde{\mathbf p}_i]_m \mathbf z_m 
\quad \Rightarrow \quad
\mathbf w_i = \sum_{m=1}^{M} [\tilde{\mathbf p}_i]_m \mathbf U^\top \mathbf z_m.
\end{align}
At this point, the nodes follow the same signal encoding in \eqref{xenc} and randomized transmission protocol described in \secref{bckr}, so that node $i$ receives the signal $\mathbf y_i$ as in \eqref{sigmod}. 
To estimate $\bar {\mathbf s}_i$ (component (a) in \eqref{didecomp}), node $i$ first computes the received sample energy as
\begin{align}
\label{fdgnhdsf}
[\mathbf r_i]_m{=}\chi_i \frac{|[\mathbf y_i]_m|^2}{(1 - p_{tx}) p_{tx} E M}, 
\ \forall m = 1, \dots, M.
\end{align}
Note that  $\mathbf r_i=\mathbf 0$ for nodes that operate as transmitters.
Compared to the baseline approach in \eqref{rim}, there is no compensation for the noise energy $N_0$, since the interference energy contribution is unknown.  
Finally, node $i$ estimates $\bar {\mathbf s}_i$ as
\begin{align}
\label{urz}
\hat {\mathbf s}_i\triangleq\mathbf U^\top \sum_{m=1}^{M} [\mathbf r_i]_m \mathbf z_m,
\end{align}
analogous to the step in \eqref{dik}, followed by a rotation back to the original frame of reference via $\mathbf U^\top$.

Note that, taking the expectation of $[\mathbf r_{i}]_m$, conditional on the frame of reference rotation $\mathbf U$, we obtain
\begin{align}
\mathbb E[[\mathbf r_{i}]_m|\mathbf U,\!\mathbf w_j,\!\forall j]{=}
\sum_{j\neq i} \Lambda_{ij}[\tilde{\mathbf p}_{j}]_m
{+}\frac{\mathbb E[\chi_i|[\mathbf n_{i}]_m|^2]}{(1{-}p_{tx})p_{tx}EM},
\label{exprim}
\end{align}
where we used the fact that the interference $\mathbf n_i$ is uncorrelated with the fading channels $h_{ij}$,
and the independence of $\chi_i$ and $\mathbf n_{i}$ from $\mathbf U$.
It then follows
$$
\mathbb E[\hat {\mathbf s}_i|\mathbf U,\!\mathbf w_j,\!\forall j]
=
\sum_{j\neq i} \Lambda_{ij}\mathbf w_j
+\sum_{m=1}^{M}\frac{\mathbb E[\chi_i|[\mathbf n_{i}]_m|^2]}{(1{-}p_{tx})p_{tx}EM}\mathbf U^\top\mathbf z_m,
$$
after replacing \eqref{exprim} into \eqref{urz}, since $\chi_i$, $\mathbf n_{i}$ are statistically independent of $\mathbf U$.
Finally, taking expectation with respect to $\mathbf  U$, and using $\mathbb E[\mathbf U]=\mathbf 0$, we obtain
\begin{align}
\label{expsi}
\mathbb E[\hat {\mathbf s}_i|\mathbf w_j,\!\forall j]
=
\sum_{j\neq i} \Lambda_{ij}\mathbf w_j,
\end{align}
so that $\hat {\mathbf s}_i$ is an unbiased estimate of $\bar {\mathbf s}_i$.

\subsubsection{Estimation of the sum of the average channel gains, $\bar\Lambda_{i}$}
Next, we address the estimation of the sum channel gains $\bar{\Lambda}_i$. To this end, each transmitting node sends a pseudo-random pilot sequence $\mathbf x_P \in \mathbb C^{n_P}$ of length $n_P \geq 2$, defined as $[\mathbf x_P]_m = \sqrt{E} e^{\j \phi_m}$, where $\phi_m \sim \mathcal U([0, 2\pi])$ are random phases, i.i.d. over $m$ and across time, but common across the network and independent of the interference signal. Note that $\mathbb E[\mathbf x_P] = \mathbf 0$ and $\mathbb E[\mathbf x_P\cdot\mathbf x_P^\h] = E \mathbf I$.

The received pilot observation at node $i$ is
\[
\mathbf y_{P,i} = \sum_{j:\chi_j = 0} h_{ij} \mathbf x_P + \mathbf n_{P,i},
\]
where $\mathbf n_{P,i}$ denotes the interference signal during pilot transmission.  
Upon receiving $\mathbf y_{P,i}$, node $i$ estimates $\bar{\Lambda}_i$ as
$$
\hat\Lambda_{i}=
\frac{\chi_i}{(1-p_{tx})p_{tx}E n_P(n_P-1)}\Big[
\frac{1}{E}|\mathbf x_P^\h\mathbf y_{P,i}|^2
-\Vert\mathbf y_{P,i}\Vert^2\Big]
$$$$
=
\frac{\chi_i}{(1-p_{tx})p_{tx}E n_P(n_P-1)}
\!\!\!\!\sum_{m,n=1:m\neq n}^{n_P}\!\!\!\!\!\!\!e^{\j(\phi_n-\phi_m)}\
[\mathbf y_{P,i}]_m[\mathbf y_{P,i}]_n^{*}.
$$
Note that $\hat{\Lambda}_i = 0$ for nodes operating as transmitters.

To compute the expectation of this estimate, consider
$$
e^{\j(\phi_n-\phi_m)}[\mathbf y_{P,i}]_m[\mathbf y_{P,i}]_n^*
=
E\Big|\sum_{j:\chi_j=0} h_{ij}\Big|^2
$$$$
+e^{\j(\phi_n-\phi_m)}[\mathbf n_{P,i}]_m[\mathbf n_{P,i}]_n^*
$$$$
+\sqrt{E}\sum_{j:\chi_j=0}(h_{ij}^*e^{-\j \phi_m}[\mathbf n_{P,i}]_m
+h_{ij}e^{\j \phi_n}[\mathbf n_{P,i}]_n^*).
$$ 
Taking the expectation with respect to the pseudo-random pilot phases and using the facts that $\mathbb E[e^{\j \phi_n}] = 0$, $\mathbb E[e^{\j (\phi_n - \phi_m)}] = 0$ for $n \neq m$, and that the interference signal is independent of the pilot phases, we obtain for $n\neq m$
$$
\mathbb E[e^{\j(\phi_n-\phi_m)}[\mathbf y_{P,i}]_m[\mathbf y_{P,i}]_n^*]
=
E\cdot\mathbb E\Big[\Big|\sum_{j:\chi_j=0} h_{ij}\Big|^2\Big]
$$$$
=
Ep_{tx}\sum_{j\neq i} \Lambda_{ij},
$$ 
where the last equality follows by taking the expectation with respect to the random transmission decisions and the Rayleigh fading coefficients.
Hence, it follows that
\begin{align}
\label{explam}
\mathbb E[\hat{\Lambda}_i]
&=
\frac{1}{n_P (n_P - 1)}
\sum_{\substack{m,n=1 \\ m \neq n}}^{n_P}
\sum_{j \neq i} \Lambda_{ij}
= \sum_{j \neq i} \Lambda_{ij}
= \bar{\Lambda}_i,
\end{align}
so that $\hat{\Lambda}_i$ is an unbiased estimate of the sum of channel gains $\bar{\Lambda}_i$.

Finally, node $i$ combines the two estimates to estimate the disagreement signal as
\begin{align}
\label{dikest}
\hat{\mathbf d}_i
&= \hat{\mathbf s}_i - \hat{\Lambda}_i \mathbf w_i
= \mathbf U^\top
\sum_{m=1}^{M} [\mathbf r_i]_m \mathbf z_m
- \hat{\Lambda}_i \mathbf w_i.
\end{align}
Using \eqref{expsi} and \eqref{explam}, it follows directly that $\hat{\mathbf d}_i$ is an unbiased estimate of the disagreement signal $\mathbf d_i$, and hence the convergence properties established in \cite{10680589} remain valid.

In the next section, we present numerical results demonstrating the performance gains of IR-NCOTA over the baseline NCOTA algorithm in the presence of interference.

\begin{figure*}
     \centering
               \hfill
          \begin{subfigure}[b]{0.32\linewidth}
        \includegraphics[width = \linewidth,trim=10 0 30 20, clip]{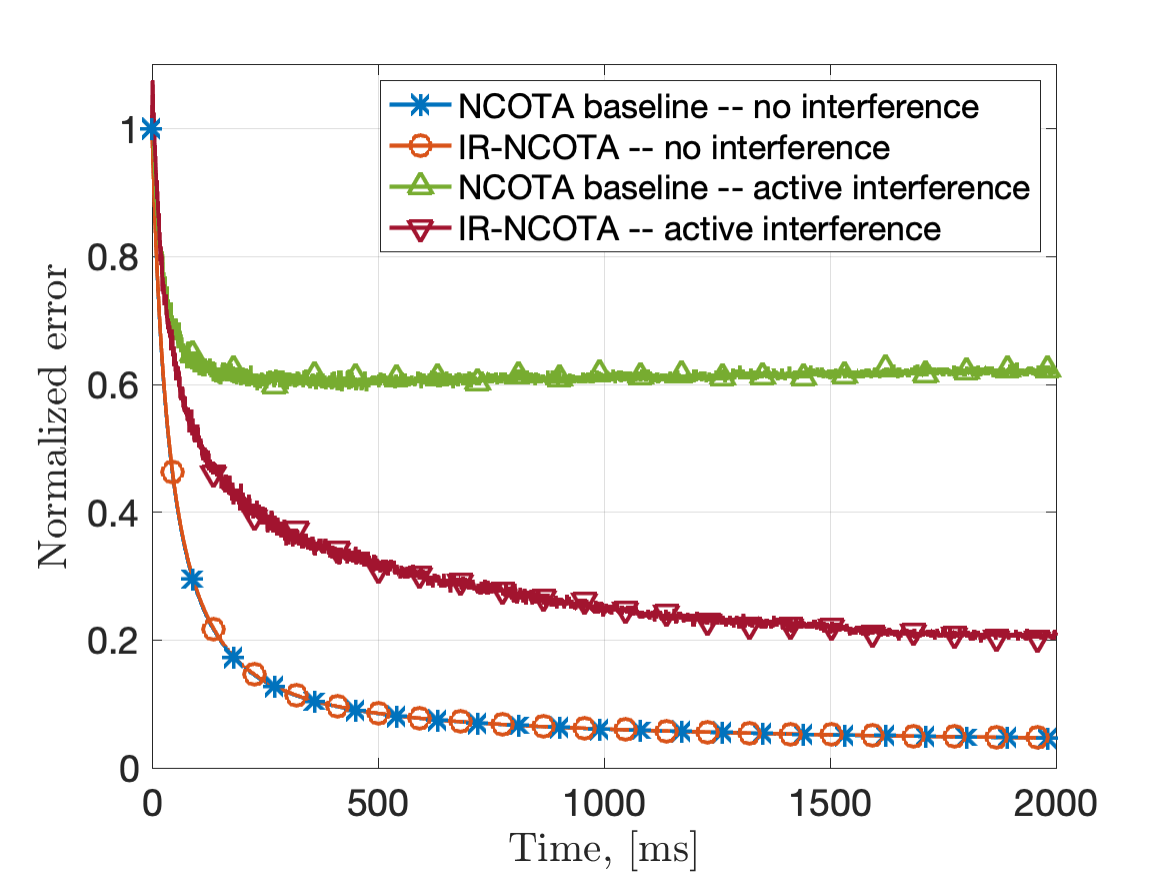}
        	    \vspace{-6mm}
	    \caption{\vspace{-3mm}} 
     \end{subfigure}
     \hfill
     \begin{subfigure}[b]{0.32\linewidth}
         \includegraphics[width = \linewidth,trim=10 0 30 20, clip]{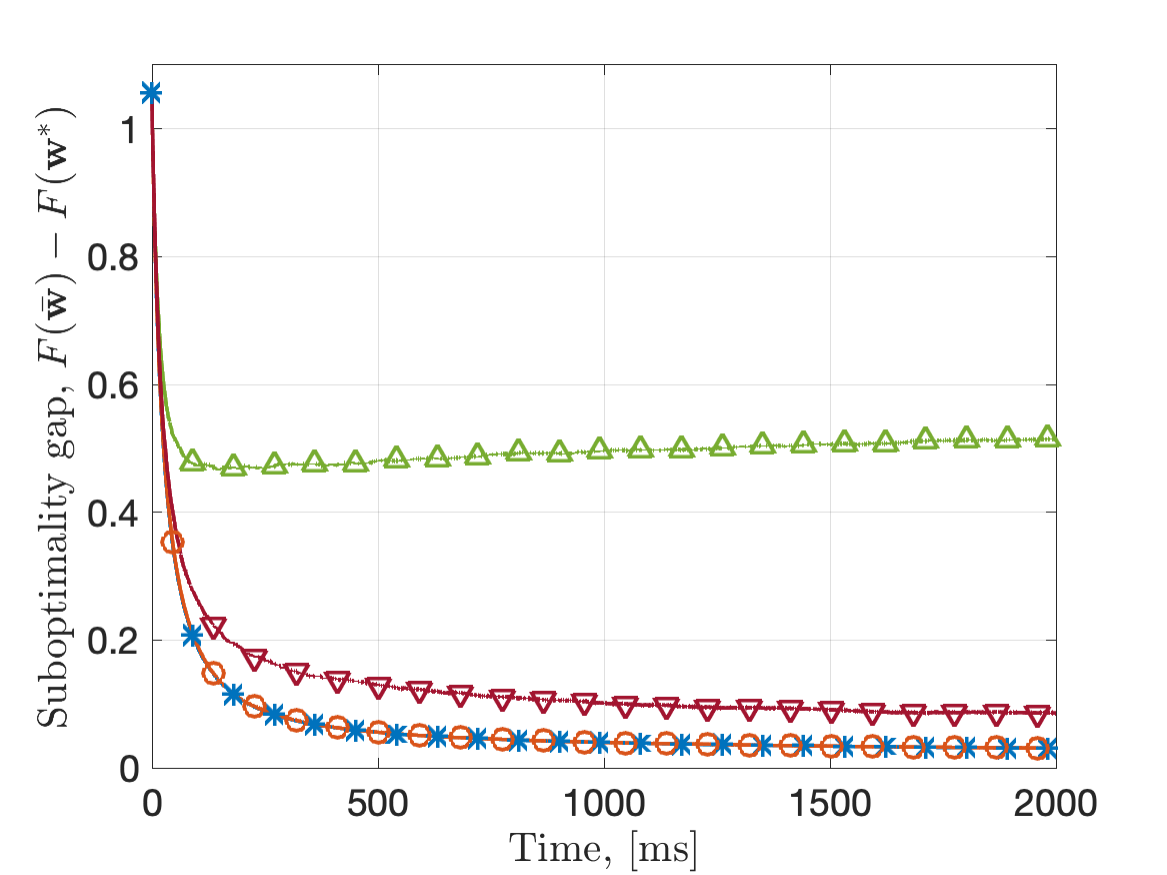}
         \vspace{-6mm}
	    \caption{\vspace{-3mm}} 
     \end{subfigure}
     \hfill
     \begin{subfigure}[b]{0.32\linewidth}
        \includegraphics[width = \linewidth,trim=10 0 30 20, clip]{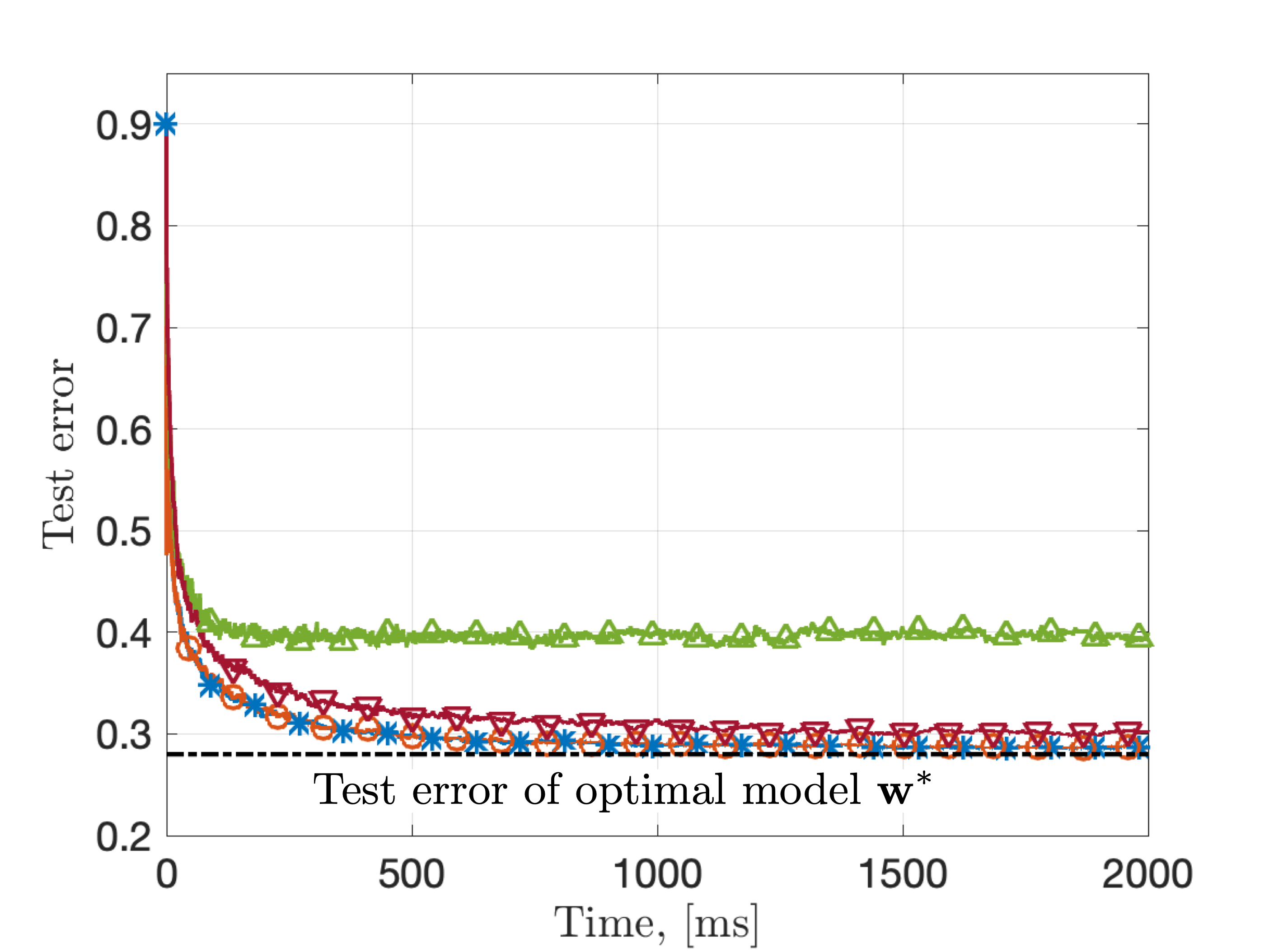}
       \vspace{-6mm}
	    \caption{\vspace{-3mm}} 
     \end{subfigure}
          \hfill
\caption{Normalized error (a), suboptimality gap (b), and test error (c), vs execution time.
Common legend shown in figure (a).
\vspace{-6mm}
 } \label{fig:SoAvsN}
\end{figure*}

\vspace{-2mm}
\section{Numerical Results}
\label{numres}
We evaluate the performance of the proposed IR-NCOTA algorithm on a classification task using the Fashion-MNIST dataset~\cite{fminst}, which contains grayscale images of fashion products from 10 classes. 

\emph{Network deployment:} $N = 200$ nodes are uniformly and independently distributed over a circular area with a radius of $2$~km. The nodes communicate over a total bandwidth of $W_{\text{tot}} = 5$~MHz at a carrier frequency of $f_c = 3$~GHz, with transmit power $P_{\text{tx}} = 20$~dBm. The receiver noise power spectral density is $N_0 = -173$~dBm/Hz.

\emph{Channel model:} Channels are modeled as Rayleigh fading, $h_{ij} \sim \mathcal{CN}(0, \Lambda_{ij})$, independently across node pairs and DGD iterations. The average channel gain $\Lambda_{ij}$ follows Friis’ free-space path-loss model, i.e., 
\[
\Lambda_{ij} = \left( \frac{\lambda}{4\pi d_{ij}} \right)^2,
\]
where $\lambda$ is the signal wavelength and $d_{ij}$ is the distance between nodes $i$ and $j$.

\emph{Data deployment:} Each node stores a local dataset of five low-resolution images belonging to a single class. Thus, 20 nodes hold data with label ‘0’, 20 with label ‘1’, and so on. Let $\ell_i \in \{0, \dots, 9\}$ denote the label associated with node $i$. Each $7 \times 7$ pixel image is transformed into a 50-dimensional feature vector $\mathbf f \in \mathbb R^{50}$ (including a bias term) and normalized such that $\|\mathbf f\| = 1$.

\emph{Optimization problem:} The learning task is formulated as a regularized logistic regression problem with loss function
\[
\phi(\ell, \mathbf f; \mathbf w)
= \frac{\mu}{2}\|\mathbf w\|^2
- \ln\!\left(
\frac{\exp\{\mathbf f^\top \mathbf w^{(\ell)}\}}
{\sum_{j=0}^{9} \exp\{\mathbf f^\top \mathbf w^{(j)}\}}
\right),
\]
where $\mathbf w^\top = [\mathbf w^{(1)\top}, \dots, \mathbf w^{(9)\top}] \in \mathbb R^d$ with $d = 450$, $\mathbf w^{(\ell)} \in \mathbb R^{50}$, $\mathbf w^{(0)} = \mathbf 0$, and $\mu = 0.001$. The local objective function at node $i$ is defined as
\[
f_i(\mathbf w) = \frac{1}{|\mathcal D_i|} \sum_{\mathbf f \in \mathcal D_i} \phi(\ell_i, \mathbf f; \mathbf w).
\]
The functions $f_i(\mathbf w)$, and consequently the global objective $F(\mathbf w)$, are strongly convex with parameter $\mu = 0.001$ and smooth with parameter $L = \mu + 2$.

\emph{Algorithms:} We compare the baseline NCOTA algorithm from~\cite{10680589} with the proposed interference-robust (IR-NCOTA) scheme under two settings:
\begin{itemize}[leftmargin=*,topsep=0pt]\setlength\itemsep{-1mm}
    \item \textbf{No interference:} in this case, there is only additive noise $\mathbf n_i \sim \mathcal{CN}(\mathbf 0, N_0 \mathbf I)$.
    \item \textbf{Active interference:} An interference source located at the center of the deployment area emits a Gaussian signal $\mathbf v\sim\mathcal{CN}(\mathbf 0, E\mathbf I)$, where $E=P_{\text{tx}}/W_{\mathrm{tot}}$ is the transmitted signal energy. This is
    received through a Rayleigh fading channel $g_i\sim\mathcal{CN}(\mathbf 0, \Gamma_i)$,
     where $\Gamma_i$ is the average channel gain between the interferer and node $i$, following Friis’ model. In this case, the total interference-plus-noise at node $i$ is $\mathbf n_i=g_i\mathbf v+\mathcal{CN}(\mathbf 0, N_0\mathbf I)$.
\end{itemize}
For the baseline NCOTA, $M=2d+1$ samples are transmitted per iteration, yielding a time-frame duration of $M/W_{\mathrm{tot}}=180.2\mu$s.
For IR-NCOTA, we generate the random unitary transformation as $\mathbf U{=}s \mathbf I$ (sign flip), where $s{=}{\pm}1$ with $\mathbb{P}(s{=}1){=}\mathbb{P}(s{=}-1){=}1/2$,
and use a pilot sequence of length $n_P=10$, corresponding to a modest pilot overhead of $\approx 1\%$, and a time-frame duration of $(M{+}n_P)/W_{\mathrm{tot}}{=}182.2\mu$s.
The transmission probability is set to $p_{tx}{=}0.34$.

\emph{Algorithm parameters:} Both algorithms use decreasing step-size sequences\footnote{We refer the interested reader to the convergence proof in~\cite{10680589} for a detailed justification of these parameter choices.}
\[
\gamma_k = \frac{\gamma_0}{(1 + k\delta)^{3/4}}, \quad
\eta_k = \frac{\eta_0}{(1 + k\delta)},
\]
for the consensus and learning steps, respectively, at DGD iteration $k$. Following the theoretical convergence analysis in~\cite{10680589}, the initial parameters are set as $\gamma_0 = 1.7 \times 10^7$, $\eta_0 = 2 / (\mu + L)$, and $\delta = \frac{5}{4\mu\eta_0}$.
Both schemes are initialized as $\mathbf w_{i}=\mathbf 0,\forall i$. The optimization set $\mathcal W$ 
has radius $r=\Vert\nabla F(\mathbf 0)\Vert/\mu$.

\emph{Evaluations and Discussion:} 
We evaluate the following performance metrics vs running time:  
(a) the \emph{normalized error} $\frac{1}{N} \sum_{i=1}^N \|\mathbf w_i - \mathbf w^*\|^2/\|\mathbf w^*\|^2$, measuring the deviation of the local models from the global optimum in~\eqref{global};  
(b) the \emph{suboptimality gap} $F\!\left(\bar{\mathbf w}\right) - F(\mathbf w^*)$ of the average model across the network, $\bar{\mathbf w}=\frac{1}{N}\sum_i \mathbf w_i$; and  
(c) the \emph{test error}, $\mathrm{TEST}\!\left(\bar{ \mathbf w}\right)$, of the averaged model, computed on a test set of 1000 samples (100 per class), with the predicted label for a feature vector $\mathbf f$ given by $\arg\max_{\ell} \mathbf w^{(\ell)\top} \mathbf f$.  

All results are averaged over 20 independent realizations of the network topology, noise, interference, transmission decisions, and—specifically for IR-NCOTA—the pseudo-random sign flips and pilot phases, as described in Secs. \ref{bckr} and \ref{sysmo}.

The results are depicted in Fig. \ref{fig:SoAvsN}.
When the interference is inactive, IR-NCOTA performs comparably to the baseline NCOTA algorithm across all metrics. All error measures exhibit a decreasing trend, with the suboptimality gap approaching zero and the test error converging to that obtained under the global optimum $\mathbf w^*$. It is worth noting that the baseline NCOTA algorithm exploits knowledge of the noise variance $N_0$ to compensate for the accumulation of noise energy, whereas the interference-robust scheme operates without such information. Nonetheless, there is no appreciable loss in performance due to the lack of knowledge of $N_0$.

When the interference is active, however, the baseline NCOTA algorithm fails to converge: all metrics reach a floor, indicating a persistent gap from the optimal solution. This behavior results from the accumulation of interference energy, which induces a drift in the learned model. In contrast, IR-NCOTA successfully mitigates this effect by making the interference contribution appear as a zero-mean process. By eliminating the drift, the error metrics under IR-NCOTA continue to decrease, although at a slower rate compared to the interference-free case due to the higher variance in the disagreement signal estimates. Overall, these results demonstrate the robustness of the proposed IR-NCOTA scheme against external interference sources.
\vspace{-2mm}
\section{Conclusions}
\label{conclu}This paper presented a novel \emph{Interference-Robust Non-Coherent Over-the-Air} (IR-NCOTA) computation scheme for decentralized optimization over wireless networks. Building upon the NCOTA framework, which enables decentralized consensus without channel state information or transmission scheduling, the proposed IR-NCOTA extends its applicability to environments affected by external interference. The core contribution lies in introducing two complementary mechanisms: a coordinated random rotation of the frame of reference and a pseudo-random pilot transmission, that jointly render the distortion introduced by the interfering signal zero-mean in expectation. This preserves the unbiasedness of the disagreement signal estimates and, consequently, the convergence guarantees of DGD. Numerical evaluations confirm that IR-NCOTA achieves comparable performance to baseline NCOTA in interference-free conditions and maintains convergence under external interference, where conventional NCOTA fails.

\bibliographystyle{IEEEtran}
\bibliography{biblio}

\end{document}